\documentclass[11pt,twoside]{book}
\usepackage{vshalo}
\usepackage{longtable}
\usepackage{amsmath,amssymb}
\usepackage{graphicx}
\usepackage{lscape}
\usepackage{psfig}
\usepackage{epsf}
\usepackage{index}
\usepackage{natbib}
\usepackage{bigdelim}
\usepackage{multirow}
\makeindex
\newindex{aut}{adx}{and}{\Large Author  Index}
\newcommand{\axindex}[1]{\index[aut]{#1}}

\begin{document}


\pagestyle{myheadings}
\setcounter{equation}{0}
\setcounter{figure}{0}
\setcounter{footnote}{0}
\setcounter{section}{0}
\setcounter{table}{0}
\setcounter{page}{19}
\markboth{Mironov et al.}{The Multicolor ``Lyra'' Photometric System 
for Variable stars and Halo Studies} 

\title{The Multicolor ``Lyra'' Photometric System for Variable stars 
and Halo Studies}
\author{Alexey V. Mironov$^1$, Andrey I. Zakharov$^1$, Mikhail E. Prokhorov$^1$, Feodor N. Nikolaev$^1$, and Maxim N. Tuchin$^1$}
\axindex{Mironov, A. V.}
\axindex{Zakharov, A. I.}
\axindex{Prokhorov, M. E.}
\axindex{Nikolaev, F. N.}
\axindex{Tuchin, M. S.}
\affil{$^1$Sternberg Astronomical Institute, Moscow, Russia} 

\begin{abstract}
The space photometric project "Lyra" is developed now in Russia.
The project purpose is determination of the photometric information and 
coordinates of the natural and artificial space objects,
from the brightest ones to 16$^m$ in visual lights. It is supposed 
to obtain the data for about 40-400 million objects from board of International 
Space Station, using an astronomical telescope with diameter of the main 
mirror of 0.5 m. The observations will be carried out in a scanning mode. 
Photometry will be obtained in 10 spectral bands. The expected uncertainty 
of magnitudes for objects of 16$^m$ in the V-band is 0.001$^m$.

The main results of experiment should become:\\
1) creation of spatial model of the Galaxy on distances to 3 kps 
from the Sun;\\
2) specification of physical parametres of stars and models of star 
evolution; \\
3) discovering of a huge number (to several millions) of variable stars 
and determination of their variability parameters.

The Sternberg Astronomical Institute of the Moscow University is the 
director of experiment and the head scientific organisation. 
Launch of the apparatus into an orbit is planning for 2013.

The central wavelengths of the 10 bands of "Lyra" photometric system 
will be at 195, 218, 270, 350, 440, 550, 700, 825, 930 and 
1000 nm. It is shown that combinations of various colour indices will 
allow to determine confidently both effective temperature and metallicity 
of stars. The presence of a 218 nm band allows to determine confidently 
interstellar extinction on stars of O – F spectral classes. 
The photometric system will make it possible to separate galo stars 
from disk stars and to define physical parameters of their atmospheres

The scanning law is that objects will be observed on the average on 
100 times each. It will give the chance to discover variables 
and to determine their characteristics.
\end{abstract}

\section{Introduction}
The space experiment "Lyra" is now being prepared in Russia.
The main goal of the experiment is to perform multicolor photometry 
for all $V<16^m$ stars.
The telescope will be mounted on the Russian segment of the 
International Space Station. 

We are planning:
\begin{itemize}
\item to make multicolor photometric catalogs for 100-400 million 
objects of $V<16^m$;\vspace{-12pt}
\item to scan repeatedly all sky during 3–5 years 
(usually 20 times per year);\vspace{-6pt}
\item to use 10 bands from 190 to 1000 nm;\vspace{-6pt}
\item to reach high precision and high accuracy of photometry 
for stars 3$^m$ -- 16$^m$;\vspace{-6pt}
\item to create an extensive uniform system of photometric 
standards;\vspace{-6pt}
\item to measure coordinates of all objects of $V<15^m$ with 
uncertainties within 1 mas and of all other objects with uncertainties within 10 mas.
\end{itemize}

Photometric system ``Lyra'' should be able to obtain basic physical 
parameters of stars and interstellar matter.

The schedule of the``Lyra'' experiment are:\\
\begin{center}
\begin{tabular}{|l|l|}
\hline
Designing and equipment manufacturing & 2008 -- 2013 \\
\hline
Performing the experiment             &              \\
Preprocessing                         & 2013 -- 2017 \\
A preliminary catalogue               &              \\
\hline
Finishing processing                  & 2017 -- 2020 \\
The final catalogue                   &              \\
\hline
\end{tabular}
\end{center}
\medskip

\section{The principles of the experiment}
The main principles of the experiment are
\begin{itemize}
\item	Scanning mode of observations;\vspace{-6pt}
\item	Time delay and integration (drift) mode;\vspace{-6pt}
\item	10-band photometry;\vspace{-6pt}
\item	Full coverage of the sky.\vspace{-6pt}
\end{itemize}

The optical scheme is classical Ritchey-Chr\'etien with a lens corrector 
from fused quartz. Main mirror diameter is 50 cm; focal distance - 3 m. 
Both mirrors shall be made from silicon carbide.

The orbital period of the ISS is 92 min. One of station axis always 
directed to Earth center.  Unmovable telescope is fastened on the board 
of ISS. The vizier axis is directed away from the Earth.  During every 
revolution round the Earth, the field of telescope's view covers a ring 
stripe on the sky. The orbit inclination is equal 51.6°. Precession 
shifts the stripe for one third of a degree in every pass of the orbit. 
Precession period is about 70 days.

If the vizier line lies in orbit plane, then, for one precession period, 
we cover a large band in the sky along the equator between -52° and +52° 
(equator mode). To reach polar regions, we must incline the axes of sight 
to the angle supplementary to orbit inclination (polar mode). 

\begin{figure}[!h]%
\centerline{\hbox{\psfig{figure=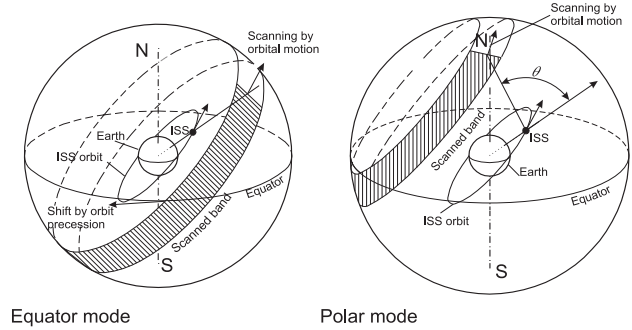,angle=0,clip=,width=9cm}}}
\caption[]{Equator and polar scanning modes} 
\label{mironov_fig1} 
\end{figure}

A choice between the two modes will depend on the position of the Sun.

The linear size of the corrected field of view will be about 78 millimeters, 
which corresponds to 1.5 degrees. 22 CCD matrices connected in pairs will be 
installed in the focal plane. The first pair (matrix 1 and 2) are uncovered 
by any filters, but they are coated with antireflective coating. 
The remaining 10 pairs of matrices will be covered by the interference 
coating so as to create 10 different photometric bands in the spectral 
range 200 - 1000 nm. During scanning the sky the stars will pass 
sequentially through all the matrices and recorded in TDI mode. 
Rocking and shaking station, distorting the trajectory of the star image 
in the focal plane will be tracked with six support matrix surrounding 
the field of main photometer.

\begin{figure}[!h]
\centerline{\hbox{\psfig{figure=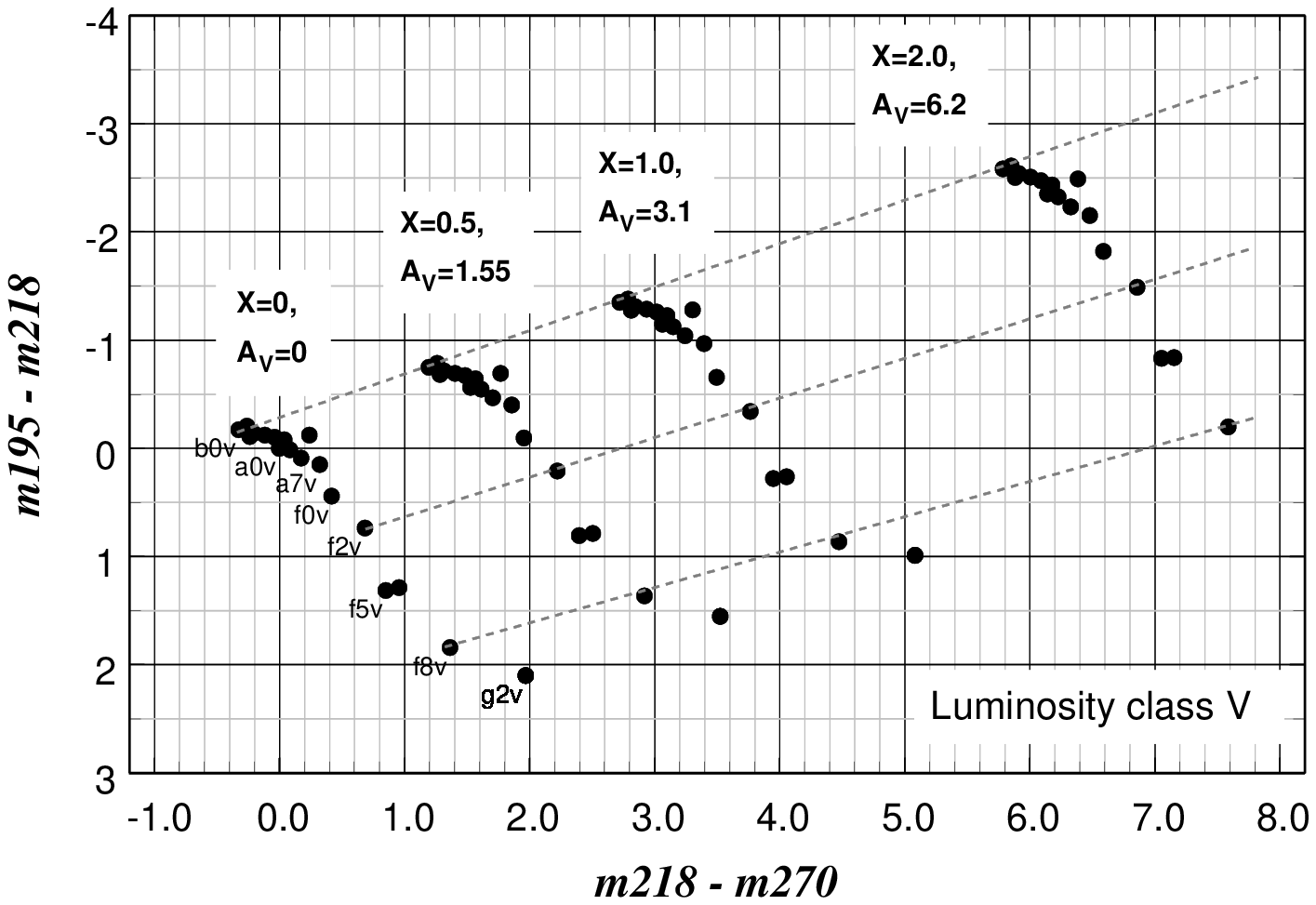,angle=0,clip=,width=9cm}}}
\caption[]{CCD matrices in the focal plane.} 
\label{mironov_fig2} 
\end{figure}

\section{``Lyra'' photometric system}
The position and half-width of the proposed photometric bands are given 
in the Table. In the same table gives the score of the brightest and most 
objects that can be measured with an error of magnitude 0.1 or 0.01.

\begin{table}
\caption{``Lyra'' photometric bands and and the expected magnitude limit 
for the mean square error of 0$^m$.01 and 0$^m$.1}
\smallskip
\begin{center}
\begin{tabular}{cccccc}
\tableline
\noalign {\smallskip} 
           &               &           &\multicolumn{3}{c}{Most faint (mag)}\\
$\lambda_0$&$\Delta\lambda$&Most bright&\multicolumn{2}{c}{for 1 observation}&for all mission \\
(nm)       &(nm)           &(mag)      &$\sigma=0.01$&$\sigma=0.1$&$\sigma=0.01$\\
\noalign{\smallskip}
\tableline
\noalign{\smallskip}
195 & 20 & 6,0 &  9,0 & 13,6 & 13,7 \\
218 & 20 & 6.0 &  8,9 &	13,6 & 13,6 \\
270 & 25 & 6,2 &  9,2 &	13,8 & 13,9 \\ 
350 & 50 & 7,6 & 10,5 &	15,1 & 15,2 \\
440 &100 & 8,8 & 12,5 &	16,9 & 17,0 \\
550 & 80 & 8,0 & 11,7 &	16,1 & 16,2 \\
700 & 80 & 6,6 & 10,7 &	15,0 & 15,1 \\
825 & 80 & 5,6 &  9,7 &	14,0 & 14,1 \\
930 & 80 & 4,3 &  8,7 &	12,9 & 12,9 \\
1000&100 & 3,3 &  7,7 &	11,9 & 12,0 \\
 
\noalign{\smallskip}
\tableline
\noalign{\smallskip}
\end{tabular}  
\end{center}

\end{table}
For the brightest objects a special mode will be applied.

\section{What can the results of the "Lyra" experiment give 
for variable stars and Halo studies?} 

\subsection{For variables stars}
Stars in sky regions near the equator will be observed, on average, 
20 times per year.  When using the polar mode, circumpolar stars will 
be observed on each pass of the orbit. Thus, each high-declination star 
will be measured several thousand times. Measuring star brightness 
simultaneously in 10 filters will permit us to reliably detect known 
variable stars and discover new ones, making use of the variable's 
brightness levels in different filters being correlated. 
The method for such discoveries was developed in the Sternberg 
Institute (Mironov et al. 2003.)
\subsection{For galactic halo}
We hope to solve at least two problems:
\begin{itemize}
\item determination of reddening (colour excesses);\hspace{-6pt}
\item determination of metallicity.\hspace{-6pt}
\end{itemize}

We have calculated several diagrams using atlases of spectral energy
distributions (SEDs) and the synthetic photometry method. As initial data we
used the atlas of empirical average SEDs by Pickles (1998) and two atlases of 
the theoretical SEDs by Lejeune et al. (1997, 1998)

On the traditional ($W-B,B-V$) diagram a reddening can be obtained 
for O - B3 spectral types only. Only for these spectra, there are no multiple 
crossings between the stellar sequences and the reddening line. 
However, stars of early spectral types usually are located near star 
formation regions where there are special conditions for interstellar matter.

\subsubsection {Determination of reddening using ultraviolet bands.}
Ultraviolet bands of the photometric system "Lyra" "195", "218" and "270" and
colour indices (m195 -- m218) and (m218 -- m270) make it possible to obtain
reliable values of reddening for stars O - F. It is shown on the figure 3.
\begin{figure}[!h]
\centerline{\hbox{\psfig{figure=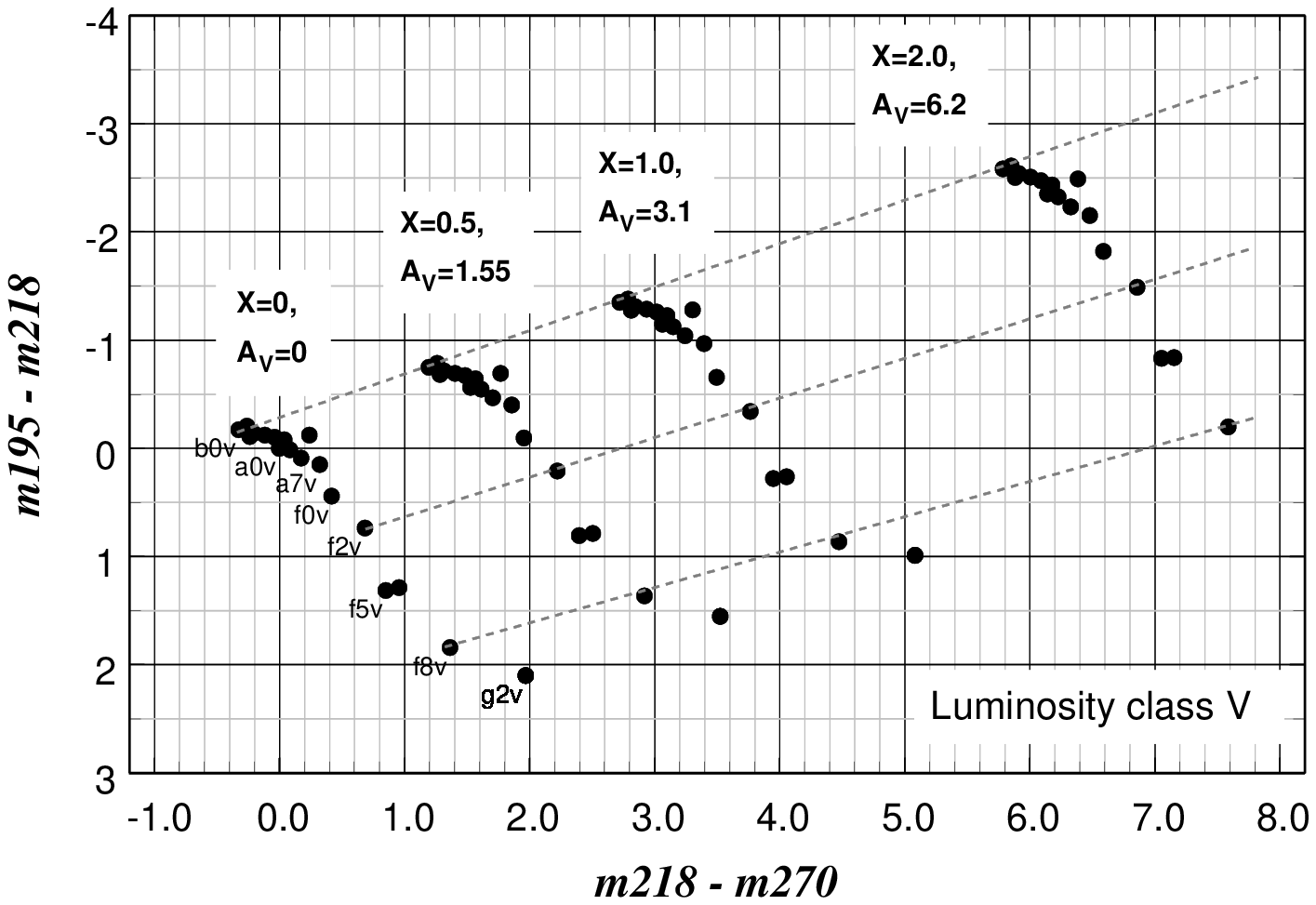,angle=0,clip=,width=12cm}}}
\caption[]{Determination of reddening for O~--~F stars} 
\label{mironov_fig3} 
\end{figure}
There are no multiple crossings between the stellar sequence and 
the reddening lines on the figure. The larger the amount of interstellar 
matter, the larger is one of the colour indices and the smaller is the 
other one. The initial data for figures 3, 4 and 5 are taken
from Pickles (1998) library.

\subsubsection {Determination of reddening using G-K stars.} 
The reddening of G -- K stars may be obtained on the two-colour diagram
(m350--m700,m700~--m930)
\begin{figure}[!h]
\centerline{\hbox{\psfig{figure=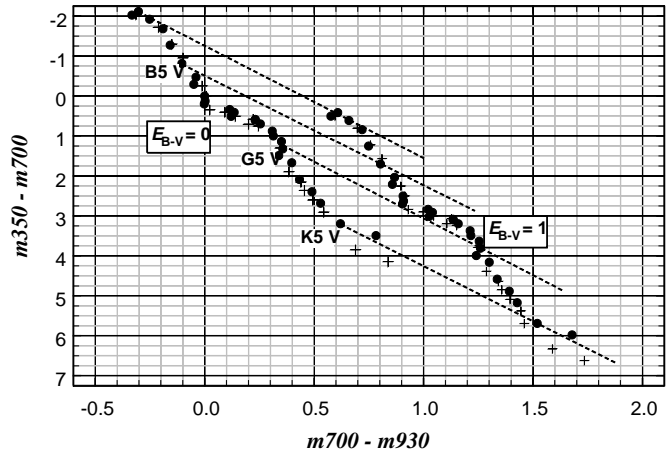,angle=0,clip=,width=12cm}}}
\caption[]{Determination of reddening for G~--~K stars. Filled circles
are main sequence stars, crosses - giants} 
\label{mironov_fig4} 
\end{figure}
On the figure 4 there are no multiple crossings 
between the stellar sequence and the reddening lines in the O-B8 and G5-K5 
spectral type ranges. 

\subsubsection {Determination of reddening using M giants.} 
Usually the slopes of temperature reddening lines and interstellar reddening 
lines on two-colours diagrams are similar. But some digrams show that for 
most red stars, the crossing angle of these lines may be not small. 
To obtain such an effect one of the used photometric bands have to be a band 
similar to V band. 
\begin{figure}[!h]
\centerline{\hbox{\psfig{figure=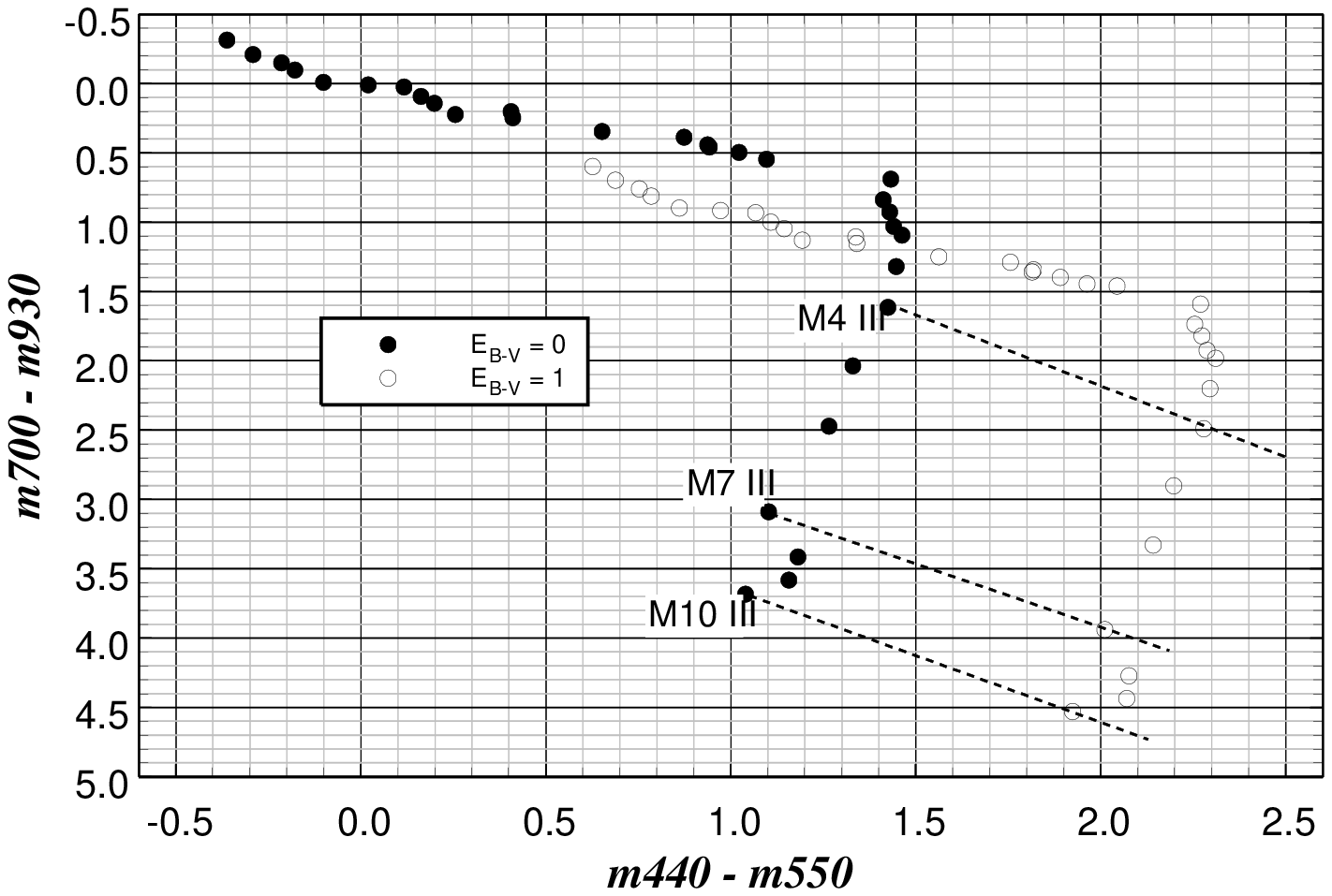,angle=0,clip=,width=12cm}}}
\caption[]{Determination of reddening using M giants} 
\label{mironov_fig5} 
\end{figure}
The two-colour diagram (m440~--~m550,m550~--m700) shown on figure 5 
allows us to determine reddening using giants M4 and later.  

\subsubsection {Differentiation of F-G-K-M stars on metallicity.} 
Figure 6 shows two-color diagram that allows us to divide the stars of 
spectral classes F - K for the metallicity.
\begin{figure}[!h]
\centerline{\hbox{\psfig{figure=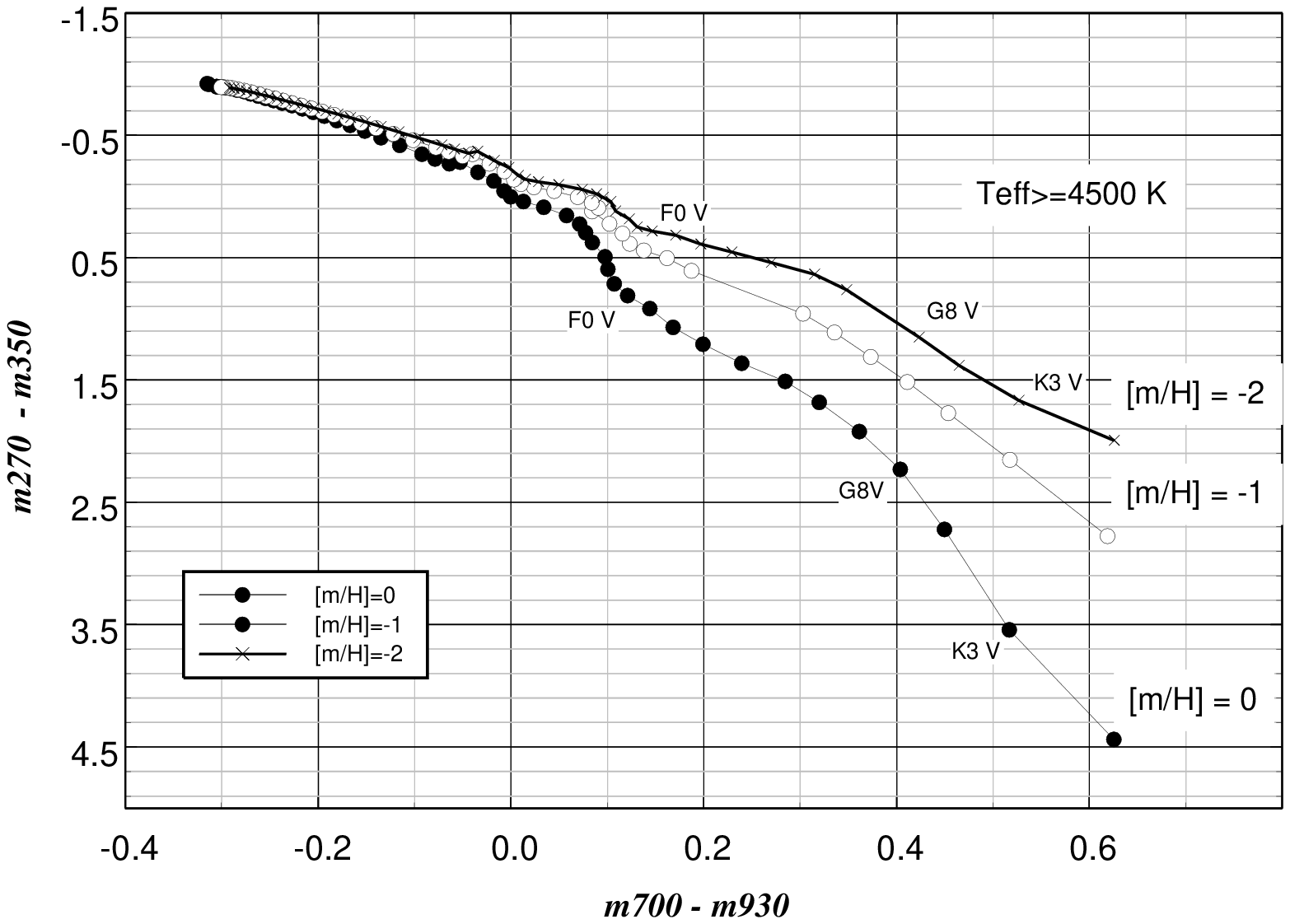,angle=0,clip=,width=12cm}}}
\caption[]{Determination of reddening using M giants} 
\label{mironov_fig6} 
\end{figure}
The initial data for the figure 6 are from Lejeune et al. (1997) and 
Lejeune et al. (1998) libraries.

We are continuing to investigate properties of the photometric system "Lyra".
 


\begin{thebibliography}{}      
\bibitem[Mironov et al.(2003)]{Mironov et al.2003} 
        Mironov A.V., Zakharov A.I., Nikolaev F.N. 2003, 
        Baltic Astronomy, 12, 589

\bibitem[Pickles A.J. (1998)]{Pickles A.J. 1998}
        Pickles A.J. 1998, Publ. Astron. Soc. Pacific, 110, 863 

\bibitem[Lejeune et al., (1997)]{Lejeune et al.1997}
	Lejeune T., Cuisinier F., Buser R. 1997,              2
  	Astron. Astrophys. Suppl. Ser. 125, 229

\bibitem[Lejeune et al., (1998)]{Lejeune et al.1998}
	Lejeune T., Cuisinier F., Buser R. 1998, 
 	Astron. Astrophys. Suppl. Ser. 130,  65
	
\end{thebibliography}
\end{document}